\documentclass[aps,preprint, showpacs]{revtex4}
\input{psfig.sty}

\usepackage{amsmath,bm,graphics}
\newcommand{\sech}{\textrm{ sech }}

\newcommand{\e}{\textrm { e}}

\newcommand{\be}{\begin{eqnarray}}
\newcommand{\ee}{\end{eqnarray}}

\begin{document}

\title{Exact Spectrum and Wave Functions of the Hyperbolic Scarf Potential
in Terms of Finite  Romanovski  Polynomials}

\author{D.\ E.\ Alvarez-Castillo, M.\ Kirchbach}

\affiliation{Instituto de F\'{\i}sica, 
          Universidad Aut\'onoma de San Luis Potos\'{\i},\\
         Av. Manuel Nava 6, San Luis Potos\'{\i}, S.L.P. 78290, M\'exico}

\vspace{0.3cm}
\begin{flushleft}
{\tt Accepted for publication by Rev.\ Mex.\ Fis.\ }
\end{flushleft}

\begin{abstract}
The Schr\"odinger equation with the hyperbolic
Scarf potential reported so far in the  literature 
is somewhat artificially manipulated into the form of the Jacobi equation 
with an imaginary argument and  parameters that are complex conjugate
to each other. Instead we here  solve the former equation anew and
make the case that it reduces straightforward to a
particular form of the generalized real hypergeometric
equation whose solutions are referred in the mathematics literature
as the finite Romanovski polynomials in reference 
to the observation that for any parameter set only a finite
number of such polynomials appear orthogonal. This is a qualitatively
new integral property that does not copy none of the features of the 
Jacobi polynomials.
In this manner the finite number of bound states within the  
hyperbolic Scarf potential is brought in correspondence to
a finite system of orthogonal polynomials of a new class.  

This work adds a new example to the circle of 
the problems on  the Schr\"odinger equation.
The techniques used by us extend the 
teachings on the Sturm-Liouville theory of ordinary differential
equations beyond their standard  presentation
in the textbooks on mathematical methods in physics.\\

\vspace{0.5cm}
\noindent  
La soluci\'on a la ecuaci\'on de Schr\"odinger con el potencial de Scarf 
hiperb\'olico reportada hasta ahora en la literatura f\'isica est\'a 
manipulada artificialmente para obtenerla en la forma de los 
polinomios de J\'acobi con 
argumentos imaginarios y par\'ametros que son complejos conjugados entre 
ellos. En lugar de eso, nosotros resolvimos la nueva ecuaci\'on obtenida y 
desarrollamos el caso en el que realmente se reduce a una forma particular 
de la ecuaci\'on hipergeom\'etrica generalizada real, cuyas soluciones se 
refieren en la literatura matemática como los polinomios finitos de 
Romanovski. La notaci\'on de finito se refiere a que, para cualquier 
par\'ametro fijo, solo un n\'umero finito de dichos polinomios son 
ortogonales. Esta es 
una nueva propiedad cualitativa de la integral que no surge como copia de 
ninguna de las características de los polinomios de Jacobi. De esta manera, 
el n\'umero finito de estados en el potencial de Scarf hiperb\'olico es 
consistente en correspondencia a un sistema finito de polinomios ortogonales 
de una nueva clase.

\end{abstract}
\pacs{02.30.Gp, 03.65.Ge, 12.60.Jv }

\maketitle

\section{Introduction}

The exactly solvable Schr\"odinger equations occupy a pole position
in quantum mechanics in so far as most of them  
relate directly to physical systems. Suffices to mention as  prominent 
examples the quantum Kepler--, or, Coulomb problem and its importance in the
description of the discrete spectrum of the
hydrogen atom \cite{TC}, the harmonic-oscillator, the Hulthen, and
the Morse potentials with their relevance to vibrational spectra
\cite{Hulten}, \cite{Franko}. Another good example is
given by the P\"oschl-Teller potential
\cite{Teller} which appears as an effective mean field in many-body
systems with  $\delta $-interactions \cite{Calogero}. 
In terms of path integrals,  the criteria for exact resolvability of the
Schr\"odinger equation is the existence of  exactly solvable
corresponding path integrals \cite{Grosche}.

There are various methods of finding the exact solutions
of the Schr\"odinger equation (SE) for the bound states, 
an issue on which we shall focus in the present work.
The traditional method, to be pursued by us here,  
consists in reducing SE by an
appropriate change of the variables to that very form of the generalized
hypergeometric equation \cite{NikUv} whose solutions are
polynomials,  the majority of them being well known.
The second method suggests to first unveil the dynamical symmetry of the
potential problem and then employ  the relevant group algebra in
order to construct the solutions as the group representation spaces
\cite{Franko_2,JPA_MG_34}.
Finally, there is also the most recent and
powerful method of the super-symmetric quantum
mechanics (SUSYQM)
which considers the special class of Schr\"odinger equations 
(in units of $\hbar =1=2m$) that
allow for a factorization according to \cite{MF}-\cite{Sukumar},
\begin{eqnarray}
\left( H(z)-e_n\right)\psi_n (z)&=&\left( 
-\frac{d^2}{dz^2} +v(z)- e_n\right)\psi_n(z)=0\, ,\nonumber\\
H(z)&=&A^+ (z)A^-(z) +e_0\, ,\nonumber\\
A^\pm(z) &=& 
\left( \pm \frac{d}{dz}+ U(z)\right)\, .
\end{eqnarray}
Here, $H(z)$ stands for the (one-dimensional) Hamiltonian, 
$U(z)$ is the so called super-potential, and 
$A^\pm (z) $ are the ladder operators connecting neighboring solutions.
The super-potential allows to recover the ground state wave function,
$\psi_{\rm gst}(z)$, as
\begin{equation}
\psi_{\rm gst}(z)\sim e^{-\int ^z U(y)dy}\, .
\label{SUSY_1}
\end{equation}
The excited states are then
built up on top of $\psi_{\rm gst}(z)$ through the repeated
action of the $A^+(z)$  operators.

\subsection{The trigonometric Scarf potential.}
The super-symmetric quantum mechanics manages a family of exactly solvable
potentials (see Refs.~\cite{Khare}--\cite{Levai}  for details)
one of which is the so called trigonometric Scarf potential (Scarf I)
\cite{Scarf_58}, here denoted by $v_t(z)$ and given by 
\begin{equation}
v_t ^{(a,b)}(z)= -a^2 +(a^2 +b^2-a\alpha )\sec^2 \alpha z -
b(2a+\alpha )\tan \alpha z
\sec \alpha z\, .
\label{Scarf_tr}
\end{equation}
It  has been used in the construction of a
periodic potential and employed in 
one-dimensional crystal models in solid state physics.

The exact solution of the Schr\"odinger equation with the
trigonometric Scarf potential (displayed in Fig.~1)   is well known 
\cite{Khare,Levai} and
given in terms of the  Jacobi
polynomials, $P_n^{\beta ,\alpha  }(x)$,  as 
\begin{eqnarray}
\psi_n(x)=\sqrt{(1-x)^\gamma (1+x)^\delta }P_n^{\gamma-\frac{1}{2},
\delta -\frac{1}{2} }(x),&\quad& x=\sin \alpha z,\nonumber\\
w^{\gamma-\frac{1}{2} ,\delta -\frac{1}{2} }(x)=
(1-x)^{\gamma-\frac{1}{2}} (1+x)^{\delta -\frac{1}{2}},
&\quad & \gamma =\frac{1}{\alpha }(a-b),
\quad \delta =\frac{1}{\alpha }(a+b)\, .
\label{Scarf_tr_sol}
\end{eqnarray}
Here, $w^{\gamma-\frac{1}{2} ,\delta-\frac{1}{2}}(x)$ 
stands for the weight function from which the Jacobi polynomials
$P_n^{\gamma-\frac{1}{2},
\delta -\frac{1}{2} }(x)$ are obtained via the Rodrigues formula.

\begin{figure}[htbp]
\centerline{\psfig{figure=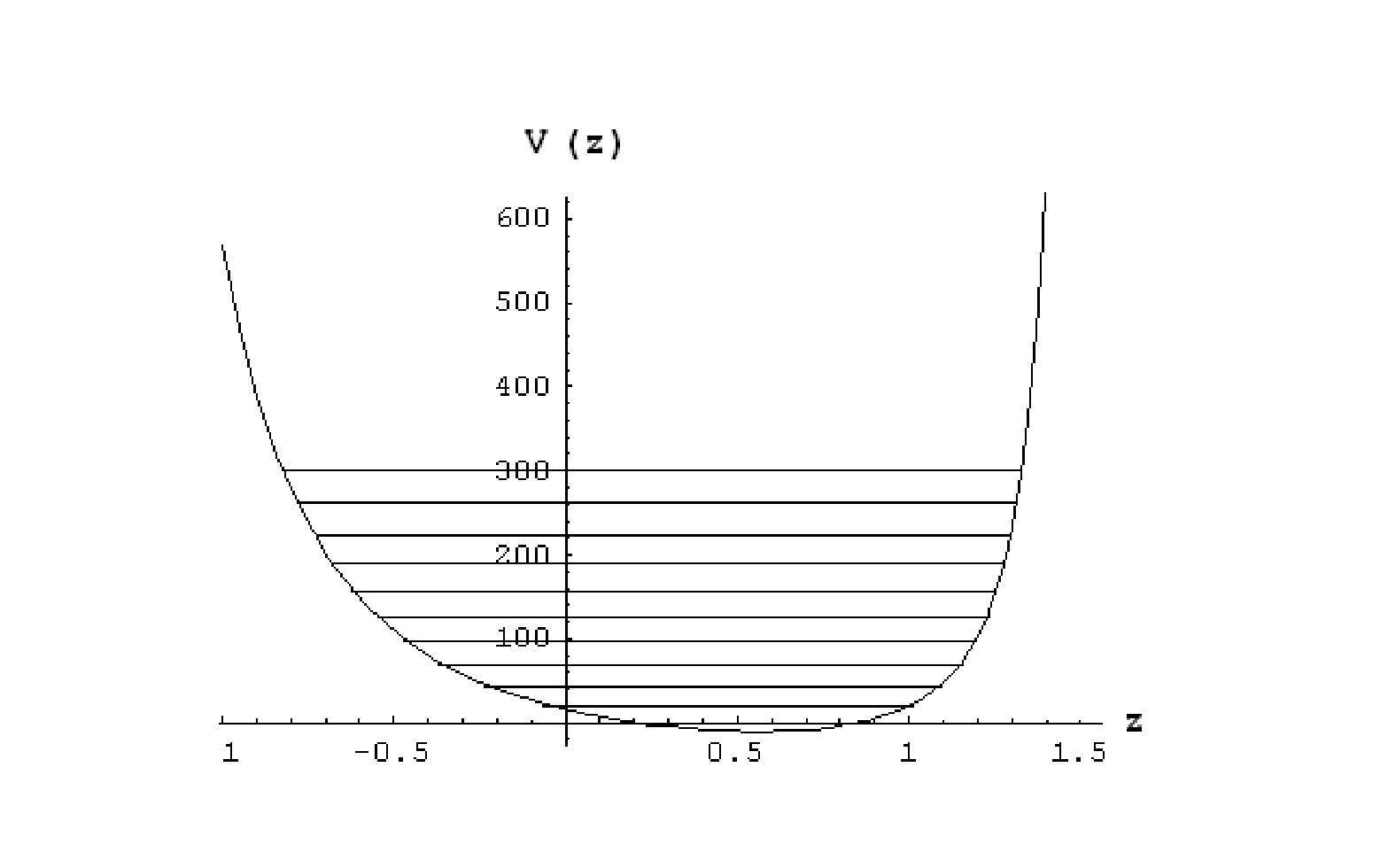,width=7cm}}\vspace{0.1cm} {\small Fig.~1.
\hspace{0.2cm}The trigonometric Scarf potential (Scarf I) for the
set of  parameters, $a=10$, $b=5$, and $\alpha =1$.
The horizontal lines represent the discrete levels.}
\end{figure}

The corresponding energy spectrum is obtained as
\begin{equation}
\epsilon_n=e_n+a^2 =(a+n\alpha)^2\, .
\label{Scarf_tr_en}
\end{equation}
\subsection{The hyperbolic Scarf potential.}
By means of the  substitutions
\begin{eqnarray}
a\longrightarrow ia\, ,&\quad& \alpha\longrightarrow -i\alpha\, ,
\quad b \longrightarrow b\, ,
\label{cmplxf}
\end{eqnarray}
Scarf I is transformed  into the so called {\it hyperbolic\/} 
Scarf potential (Scarf II), here
denoted by $v^{(a,b)}_h(z)$ and displayed in Fig.~2,
\begin{equation}
v^{(a,b)}_h(z)=a^2 +(b^2-a^2-a\alpha)\sech^2\alpha z
+b(2a+\alpha)\sech \alpha z\tanh \alpha  z\, .
\label{Scarf_hip}
\end{equation}
The latter potential has also been found independently 
within the framework of super-symmetric quantum mechanics
while exploring the super-potential ~\cite{Khare},\cite{Levai},\cite{Bagchi} 
\begin{equation}
U (z)=a\tanh \alpha z +b\sech\alpha z\,.
\label{Scarfh_superpot}
\end{equation}
Upon the above substitutions and in taking $\alpha =1$ for simplicity
the energy changes to
\begin{equation}
\epsilon_n=e_n-a^2=-(a-n)^2\, , \quad n=0,1,2,...< a\, .
\label{enrg_hip}
\end{equation}

\begin{figure}[htbp]
\centerline{\psfig{figure=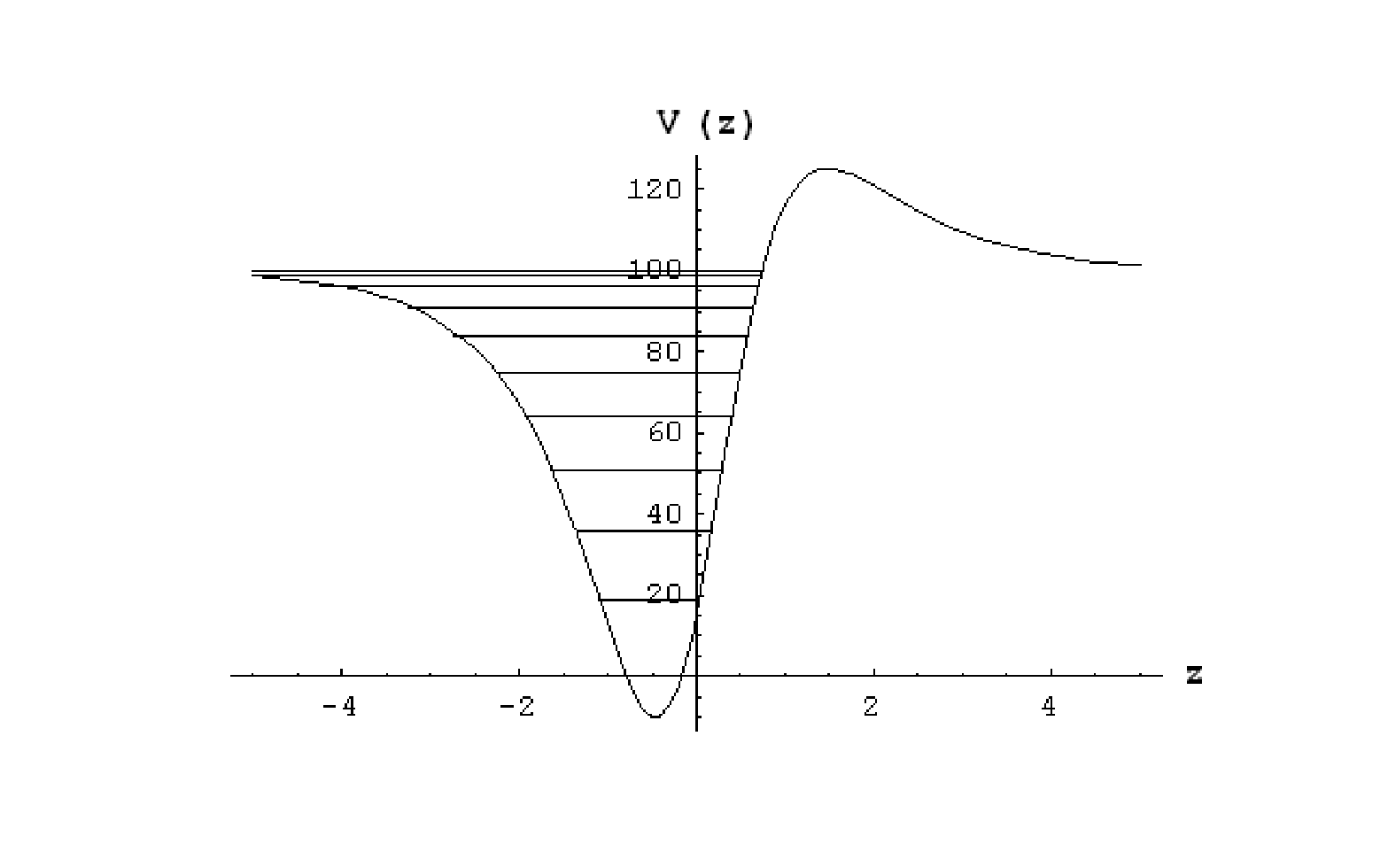,width=7cm}}
\vspace{0.1cm}{\small Fig.~2.
\hspace{0.2cm}The hyperbolic Scarf potential (Scarf II) for the 
set of parameters,  $a=10$, $b=5$, and $\alpha =1$.
The horizontal lines represent the energies, 
$e_n$, of the bound states.}
\end{figure}
It is important to notice that while the trigonometric Scarf potential 
allows for an infinite number of bound states, the number of discrete
levels within the hyperbolic Scarf potential is finite,
a difference that will be explained in section III below. 
Yet, the most violent changes seem to be suffered by the Jacobi 
weight function in eq.~(\ref{Scarf_tr_sol}) and are due to the opening
of the finite interval $\lbrack -1,+1\rbrack $ toward infinity,
\begin{equation}
x=\sin \alpha z \in \lbrack -1,1\rbrack
 \longrightarrow x=\sinh \alpha z \in \lbrack -\infty , +\infty \rbrack\, .
\end{equation}
In this case, the wave functions become
\cite{Khare},\cite{nicolae}, \cite{Hidalgo},
\begin{eqnarray}
\psi_n(-ix)
&=&  (1+x^2)^{ -\frac{a}{2} }
e^{- b \tan^{-1} x}
c_n P_n^{\eta ^\ast  , \eta}(-ix)\,,
\quad \eta  = ib -a -\frac{1}{2} \, .
\label{cmplx_Jac_x}
\end{eqnarray}
Here, $c_n$ is some state dependent complex phase to be fixed later on.
The latter equation gives the impression that the exact solutions of the 
hyperbolic Scarf potential rely exclusively upon those peculiar
Jacobi polynomials  with imaginary arguments and complex indices.
We here draw attention to the fact that this needs not be so. 

\subsection{The goal.}

\begin{quote}
The goal of this work is to solve the Schr\"odinger equation 
with the hyperbolic Scarf potential anew and to  make the case that 
it reduces in a straightforward manner to a particular form of the 
generalized real hypergeometric equation whose solutions 
are given by a finite set of real orthogonal polynomials.
In this manner, the finite number of bound states within the
hyperbolic Scarf potential is brought in correspondence
with a finite system of orthogonal polynomials of a new class.  
\end{quote}
These polynomials have been discovered in 1884 by 
the English mathematician Sir Edward John Routh
\cite{Routh} and rediscovered 45 years later by the Russian
mathematician Vsevolod Ivanovich Romanovski in 1929 
\cite{Romanovski} within the context of probability distributions. 
 Though they have been studied on few  occasions in the 
current mathematical literature where they are termed to as  
``finite Romanovski'' \cite{Zarzo}--\cite{Neretin},
or, ``Romanovski-Pseudo-Jacobi'' polynomials \cite{Lesky}, they 
have been completely ignored by
the textbooks on mathematical methods in physics,
 and surprisingly enough, by the standard mathematics textbooks 
as well \cite{NikUv}, \cite{MMF}-\cite{ism}.
The notion ``finite'' refers to the observation that for any given set of 
parameters (i.e. in any potential),
only a finite number of polynomials appear orthogonal.

The  Romanovski polynomials happen to be equal
(up to a phase factor) to Jacobi polynomials with 
imaginary arguments and parameters that are complex 
conjugate to each other, much like the 
$\sinh z=i\sin i z$ relationship. 
Although one may (but has not to) deduce the 
local characteristics of the latter such as 
generating function and recurrence relations from those of the former,
the finite orthogonality theorem is qualitatively new. 
It does not copy none of the properties of the Jacobi polynomials
but requires an independent proof.

Our work adds a new example to the circle of 
the typical quantum mechanical problems \cite{Fluegge}.
The techniques used by us here extend  the 
study of the Sturm-Liouville theory of ordinary differential
equations beyond that of the usual textbooks.  

A final comment on the 
importance of the potential in eq.~(\ref{Scarf_hip}).
 The hyperbolic Scarf potential
finds various applications in physics ranging from electrodynamics
and solid state physics to particle theory.
In solid state physics Scarf II is used in the construction
of more realistic periodic potentials in crystals \cite{Kusnezov} than
those built from the trigonometric Scarf potential. 
In electrodynamics Scarf II appears in a class of problems with 
non-central potentials (see section IV ).
In particle physics Scarf II finds application in studies of
the non-perturbative sector of gauge theories by means of 
toy models such as the scalar field theory in 
(1+1) space-time dimensions. Here, one encounters  the so called 
``kink -like'' solutions which are no more but the
static solitons. The spatial derivative of the
kink-like solution is  viewed as the ground state wave function of an 
appropriately constructed Schr\"odinger equation 
which is then employed in the calculation of the quantum corrections 
to first order.
In Ref.~\cite{Hidalgo} it was shown that specifically Scarf II is amenable
to a stable renormalizable scalar field theory.

The paper is organized as follows. In the next section we first
highlight in brief the basics of the generalized hypergeometric
equation and then relate it to the
Schr\"odinger equation with the hyperbolic Scarf potential.
The solutions are obtained  in terms of finite Romanovski polynomials
and are presented in section III. Section IV is devoted to
the disguise of the Romanovski polynomials as 
non--spherical angular functions. 
The paper ends with a brief summary.

\section{Master formulas for the polynomial solutions to the
 generalized hypergeometric equation. }

All classical orthogonal polynomials appear as solutions of the so called
generalized hypergeometric equation (the presentation in this section 
closely follows Ref.~\cite{Koepf})
\begin{eqnarray}
\sigma(x)y_n^{\prime\prime}(x)+\tau (x)y_n^\prime (x)-\lambda_ny_n(x)&=&0\, ,
\label{hyperg_EQ}\\
\sigma(x)=a x^2+bx+c,\quad \tau(x)=xd +e\,, &\quad&
\lambda_n=n(n-1)a+nd\,  . 
\end{eqnarray}
There are various methods for finding the solution, here denoted by
\begin{eqnarray}
y_n(x)&\equiv &P_n\left(
\begin{array}{ccc}
d&&e\\
a&b&c
\end{array}{\Bigg|}x \right)\, , 
\end{eqnarray}
with the symbol  $P_n\left(
\begin{array}{ccc}
d&&e\\
a&b&c
\end{array}{\Bigg|}x \right)$ in which 
the equation parameters have been made explicit 
standing for a polynomial of degree $n$, $\lambda_n$ 
being the eigenvalue parameter, and $ n=0,1,2,...$.
In Ref.~\cite{Koepf} a master formula for the (monic, ${\bar P}_{n}$),
polynomial solutions has been derived by Koepf and Masjed-Jamei,
according to them  one finds
\begin{eqnarray}
&&{\bar P}_n\left(
\begin{array}{ccc}
d&&e\\
a&b&c
\end{array}{\Bigg|}x \right)
=\sum_{k=0}^{n}
\left(
\begin{array}{c}
n\\
k
\end{array}
\right) G_k^{(n)}(a,b,c,d,e)x^k\, ,\nonumber\\
&&G_k^{(n)}=
\left( 
\frac{2a}{b+\sqrt{b^2-4ac}}
\right)^{n}\, \, 
_2F_1
\left(
\begin{array}{cc}
(k-n),&\left( 
\frac{2ae -bd}{2a\sqrt{b^2-4ac}}+1-\frac{d}{2a}-n\right) \\
2-\frac{d}{a} -2n&
\end{array}
{\Bigg|}
\frac{
2\sqrt{b^2-4ac}}
{b+\sqrt{b^2-4ac}}
\right).
\nonumber\\
\label{monic_master}
\end{eqnarray}
Though the formal proof of this relation is bit lengthy, its verification
with symbolic mathematical softwares like Maple is straightforward.
The $a=0$ case is treated as the $a\longrightarrow 0$ limit of
eq.~(\ref{monic_master}) and leads to $_2F_0$ 
in place of $\, _{2}F_{1}$.
Notice that  $G_k^{(n)}$  are not normalized.
On the other side, eq.~(\ref{hyperg_EQ}) 
can be treated alternatively as described in
the textbook by Nikiforov and Uvarov in Ref.~\cite{NikUv} 
where it is cast into a self-adjoint form and 
its weight function, $w(x)$, satisfies the so called Pearson differential
equation,
\begin{equation}
\frac{\partial }{dx}\left( \sigma (x) w(x)\right)=\tau (x)w(x)\, .
\label{Pearson}
\end{equation}
The Pearson equation is solved by
\begin{equation}
w(x)\equiv {\mathcal W}
\left(
\begin{array}{ccc}
d&&e\\
a&b&c
\end{array}
{\Bigg|}x
\right)
=\exp \left(
\int \frac{(d-2a)x+(e-b)}{ax^2+bx+c}dx
\right)\, ,
\label{weight_general}
\end{equation}
and shows how one can calculate any weight function 
associated with any parameter set of interest
(we again used a symbol  for the weight function  
that makes explicit the parameters of the equation). 
The corresponding  polynomials are now classified according to 
the weight function and are built up from the Rodrigues representation as
\begin{eqnarray}
 P_n\left(
\begin{array}{ccc}
d&&e\\
a&b&c
\end{array}{\Bigg|}x \right)&=&
\Pi_{k=1}^{k=n}(d +(n+k-2)a){\bar P}_n\left(
\begin{array}{ccc}
d&&e\\
a&b&c
\end{array}{\Bigg|}x \right)=
\frac{1}{
{\mathcal W}
\left(
\begin{array}{ccc}
d&&e\\
a&b&c
\end{array}
{\Bigg|}x
\right)
}\nonumber\\
&\times &
\frac{
d^n
}{dx^n}\left( (ax^2+bx+c)^n
{\mathcal W}
\left(
\begin{array}{ccc}
d&&e\\
a&b&c
\end{array}
{\Bigg|}x
\right)\right)\, .\nonumber\\
\label{Rodrigues}
\end{eqnarray}
The master formulas in the respective
eqs.~(\ref{monic_master}), and (\ref{Rodrigues}) allow for the 
construction of all the polynomial solutions to
the generalized hypergeometric equation.
One identifies as special cases the canonical parameterizations
known as 
\begin{itemize}
\item   the
Jacobi polynomials with $a=-1$, $b=0$, $c=1$, 
$d=-\gamma -\delta  -2$, and $e=-\gamma +\delta$,
\item the Laguerre polynomials with
$a=0$, $b=1$, $c=0$, $d=-1$, and $e=\alpha +1$,
\item  the  Hermite polynomials with  $a=b=0$, $c=1$, $d=-2$, 
and $e=0$,
\item the Romanovski polynomials  with
$a=1$, $b=0$, $c=1$, $d=2(1-p)$, and $e=q$ with $p>0$,
\item the Bessel polynomials with $a=1$, $b=0$, $c=0$, $d=\alpha +2$, 
and $e=\beta $.
\end{itemize}
All other parameterizations can be reduced to one of 
the above five by an appropriate shift of the variables.
The  first three  polynomials are the only ones that are 
traditionally  presented in the standard textbooks on mathematical 
methods in physics such like \cite{MMF}--\cite{ism},
while the fourth and fifth seem to have escaped due attention.
Notice, the Legendre, Gegenbauer,  and Chebychev polynomials
appear as particular cases of the Jacobi polynomials.
The Bessel polynomials are not orthogonal in the conventional
sense, i.e. within a real interval, and will be left out of consideration.

Some of the properties of the fourth polynomials have been 
studied in the specialized mathematics literature such as 
Refs.~\cite{Zarzo},\cite{Mohamed}, \cite{Neretin}.
Their weight function is calculated from eq.~(\ref{weight_general}) as
\begin{equation}
w^{(p ,q)}(x)=(x^2+1)^{-p}e^{ q\tan^{-1}x}\, .
\label{wafu_Rom}
\end{equation}
This weight function has first been reported by Routh \cite{Routh}, and
independently Romanovski \cite{Romanovski}.
The polynomials associated with  eq.~(\ref{wafu_Rom})
are named after Romanovski and will be 
denoted by $R_m^{(p,q)}(x)$. They
have  non-trivial orthogonality properties
over the infinite interval
 $[-\infty, +\infty ]$. 
Indeed, as long as the weight function decreases as 
$x^{-2p}$, hence  integrals of the type
\begin{equation}
\int_{-\infty}^{+\infty}w^{(p,q)}(x)
R_m^{(p,q)}(x) R_{m^\prime}^{(p,q)}(x) dx
\label{orth_int}
\end{equation}
are convergent only if
\begin{equation}
m+m^\prime < 2p-1\, ,
\label{orth_cond}
\end{equation}
meaning that only a finite number of Romanovski polynomials are orthogonal.
This is the reason for the term  ``finite''Romanovski polynomials
(details are given Ref.~\cite{Raposo}).
The differential equation satisfied by the Romanovski polynomials reads
as
\begin{equation}
(1+x^2)\frac{d ^2 R_n^{(p,q)}(x)}{d^2 x}
+\left( 2(-p+1)x +q\right)\frac{d R_n^{(p,q)}(x)}{dx}
-(n(n-1)+2n(1-p))R_n^{(p,q)}(x)=0\, .
\label{Rom_pol}
\end{equation}
In the next section we shall show that the Schr\"odinger
equation with the hyperbolic Scarf potential reduces precisely to that very 
eq.~(\ref{Rom_pol}).

\subsection{The real polynomial equation associated with 
the hyperbolic Scarf potential.}
 
The Schr\"odinger equation for the potential of interest when rewritten in
a new  variable, $x$, introduced via an appropriate point canonical 
transformation \cite{Wipf}, \cite{De}, taken by us as
$z=f(x)=\sinh^{-1} x$, is obtained as:
\begin{equation}
(1+x^2)\frac{d^2g_n(x)}{dx^2}+x\frac{dg_n(x)}{dx}
+\left( \frac{-b^2 +a(a+1)}{1+x^2} 
-\frac{b(2a+1)}{1+x^2}x
+\epsilon_n  \right)g_n(x)=0\,,
\label{Schr_bzero}
\end{equation}
with $g_n(x)=\psi_n(\sinh^{-1} x )$, and $\epsilon_n=e_n-a^2$.
Equation~(\ref{wafu_Rom})
suggests the following substitution in eq.~(\ref{Schr_bzero})
\begin{eqnarray}
g_n(x)&=&(1+x^2)^{\frac{\beta }{2}}e^{-\frac{\alpha}{2}\tan^{-1} x}
D^{(\beta ,\alpha )}_n(x)\, , \quad x=\sinh z\,, \quad -\infty <x<+\infty.
\label{azero_1}
\end{eqnarray}
In effect, eq.~(\ref{Schr_bzero}) reduces to the following equation
for $D^{(\beta,\alpha )}_n(x)$,
\begin{eqnarray}
(1&+&x^2)
\frac{
d^2D^{(\beta ,\alpha )}_n(x)}{dx^2}
+( (2\beta +1)x -\alpha )\frac{dD^{(\beta ,\alpha )}_n(x)}{dx}\nonumber\\
&+&\left(\beta^2 +\epsilon_n +
\frac{(a+a^2+\beta -\beta^2-b^2 +\frac{\alpha^2}{4})
+x(-b -2ab +\frac{\alpha}{2}-\alpha\beta )
}{1+x^2} 
\right)D^{(\beta,\alpha )}_n(x)=0\, .
\nonumber\\
\label{azero_2}
\end{eqnarray}
If the potential  equation~(\ref{azero_2})  is to coincide with 
the Romanovski equation~(\ref{Rom_pol}) then 
\begin{itemize}
\item first  the coefficient in front of
the $1/(x^2+1)$ term in (\ref{azero_2}) has to vanish,
\item the coefficients in front of the first derivatives have to be equal,
i.e. $2(-p+1)+q=(2\beta +1)x -\alpha $,
\item the eigenvalue constants should be equal too, i.e.
$\epsilon_n +\beta^2 =- n((n-1) +2(1-p))$.
\end{itemize}
The first condition restricts  the parameters
of the $D_n^{(\beta ,\alpha )}(x)$ polynomials to
\begin{eqnarray}
a+a^2-b^2+\frac{\alpha^2}{4} +\beta -\beta^2&=&0\, ,
\label{1s_cond}\\
-b-2ab+\frac{\alpha}{2}-\alpha\beta &=&0\, .
\label{2nd_cond}
\end{eqnarray}
Solving the  equations (\ref{1s_cond}), (\ref{2nd_cond}), for 
$\alpha$ and $\beta$ results in
\begin{eqnarray}
\beta =-a\, , &\quad& \alpha =2b\, .
\label{azero_4}
\end{eqnarray}
The second condition relates the parameters $\alpha$ and $\beta $
to $p$, and $q$, and   amounts to
\begin{equation}
\beta=-a=-p+\frac{1}{2}, \quad -\alpha=q=-2b.
\label{p_q_par}
\end{equation}
Finally, the third restriction
leads to a condition that fixes the Scarf II energy spectrum as 
\begin{equation}
\epsilon_n=-(a -n)^2\,.
\label{azero_3}
\end{equation}
In this way, the  polynomials
that enter the solution of the Schr\"odinger equation will be
\begin{equation}
D_n^{(\beta =-a, \alpha  =2b)}(x)
\equiv R_n^{\left(p=a+\frac{1}{2}, q=-2b\right)}(x).
\label{bong}
\end{equation}
They are obtained by means of the Rodrigues formula from the
weight function $w^{(a+\frac{1}{2}, -2b )}(x)$ as
\begin{eqnarray}
R_m^{(a+\frac{1}{2}, -2b)}(x)&=&
\frac{1}{w^{(a+\frac{1}{2}, -2b)}(x)}\frac{d^m}{dx^m}
(1+x^2)^m {w^{(a+\frac{1}{2}, -2b)}(x)}\,,\nonumber\\
{w^{(a+\frac{1}{2}, -2b)}(x)}&=&(1+x^2)^{-a-\frac{1}{2}}
e^{-2b \tan^{-1}x}\, .
\label{Rod_Rom}
\end{eqnarray}
As a result, the wave function of the $n$th level, $\psi_n$,  
takes the form
\begin{eqnarray}
\psi_n(z=\sinh^{-1}x ):\stackrel{\mathrm{def}}{=}g_n(x)&=&\frac{1}{
\sqrt{\frac{d\sinh^{-1}x }{dx}} } \sqrt{
(1+x^2)^{-\left( a+\frac{1}{2}\right) }
e^{-2b\tan^{-1} x}} 
R_n^{\left(a+\frac{1}{2},-2b\right)}(x), \nonumber\\
d\sinh^{-1}x  &=&\frac{1}{\sqrt{1+x^2}}dx\,.
\label{wfu_x}
\end{eqnarray}
The orthogonality integral of the Schr\"odinger wave functions gives rise
to the following orthogonality integral of the Romanovski polynomials,
\begin{equation}
\int_{-\infty}^{+\infty} \psi_n(z)\psi_{n^\prime}( z)d z=
\int_{-\infty}^{+\infty}
(1+x^2)^{-\left( a+\frac{1}{2}\right) }
e^{-2b\tan^{-1}x}R_n^{(a+\frac{1}{2},-2b)}(x)
R_{n^\prime}^{(a+\frac{1}{2},-2b)}(x)dx\, ,
\label{orth_Schr_wafu}
\end{equation}
which coincides in form with the integral in eq.~(\ref{orth_int}) and
is convergent for $n<a$. 
That only a finite number of Romanovski polynomials are orthogonal,
is reflected by the finite number of bound states within the potential of 
interest, a number that depends on the potential parameters
in accord with eq.~(\ref{orth_cond}).

As to the complete Scarf II spectrum, it  has been constructed
in Ref.~\cite{JPA_MG_34} within the  dynamical symmetry approach
\cite{Franko_2}. 
There, the Scarf II potential  has been found to possess 
$SU(1,1)$ as a symmetry group. 
The bound states have been assigned to the discrete unitary irreducible 
representations of $SU(1,1)$. 
The scattering and resonant
states (they are beyond the scope of the present study)
have been related to the continuous unitary
 and the non-unitary representations of
$SU(1,1)$, respectively.

A comment is in place on the relation between the Romanovski polynomials 
and the Jacobi polynomials of imaginary arguments and parameters that
are complex conjugate to each other. Recall 
the real Jacobi equation,
\begin{equation}
(1-x^2)\frac{d^2P_n^{\gamma,\delta }(x)}{dx^2}
+(\gamma -\delta  -(\gamma +\delta +2)x)\frac{dP_n^{\gamma,\delta}(x)}{dx}
-n(n+\gamma +\delta +1)P_n^{\gamma, \delta }(x)=0\, .
\label{Jacobi}
\end{equation}
As mentioned above, the real Jacobi polynomials are orthogonal within
the $[-1,1]$ interval with respect to the weight-function
in eq.~(\ref{Scarf_tr_sol}).
Transforming to complex argument, $x\to ix$, and  parameters,
$\gamma=\delta^\ast=c+id$,
eq.~(\ref{Jacobi}) transforms into
\begin{eqnarray}
(1+x^2)\frac{d^2P_n^{c+id,c-id }(ix)}{dx^2}&+&
(-2d  +2(c +1)x)\frac{dP_n^{c+id,c-id }(ix)}{dx}\nonumber\\
&+&n(n+2c +1)P_n^{c+id , c-id  }(ix)=0\, .
\label{Jacobi_cplx}
\end{eqnarray}
Correspondingly, the weight function turns to be
\begin{equation}
w^{c+id, c-id }(ix)=(1+x^2)^c e^{-2d\tan ^{-1} x}\, ,
\label{wghtC}
\end{equation}
and it coincides with the weight function of the Romanovski polynomials
in eq.~(\ref{wafu_Rom}) upon the identifications $c=-p$, and $q=-2d$.
This means that $P_n^{c+id,c-id }(ix)$ will differ from the 
Romanovski polynomials by a phase factor found as $i^n$ in 
Ref.~\cite{Levai_PL}, among others,
\begin{equation}
i^n P_n\left(
\begin{array}{ccc}
2(1-p)&& iq\\
-1&0&1
\end{array}{\Bigg|}ix \right)=
 P_n\left(
\begin{array}{ccc}
2(1-p)&&q\\
1&0&1
\end{array}{\Bigg|}x \right)\ .
\label{Rom_Jac}
\end{equation}
Because of this relationship the Romanovski polynomials have been termed to
as ``Romanovski-Pseudo-Jacobi'' by Lesky ~\cite{Lesky}. 
The  relationship in eq.~(\ref{Rom_Jac}) tells that  the $R^{(p,q)}_n(x)$ 
properties  translate into those of
 $P^{-p -i\frac{q}{2}, -p +i\frac{q}{2}}(ix)$ 
and visa versa, and that it is a matter of convenience to prefer 
the one polynomial  over the other.
When it comes up to recurrence relations, generating functions etc.
it is perhaps more convenient to favor the Jacobi polynomials, 
though the generating function of the
Romanovski polynomials is also equally well calculated directly
from the corresponding Taylor series expansion \cite{Raposo}.
However, concerning the orthogonality integrals, the advantage
is clearly on the side of the real Romanovski
polynomials. This is so because the complex Jacobi polynomials are 
known for their highly non-trivial orthogonality properties which depend 
on the interplay between integration contour and  parameter values 
\cite{Jacobi-c}.
{}For this reason, in  random matrix theory \cite{Witte} the problem on the 
gap probabilities in the spectrum of the circular Jacobi ensemble
is treated in terms of the Cauchy random ensemble,
a venue that heads one again to the Romanovski polynomials 
(notice that for $p=1, q=0$ the weight function in eq.~(\ref{wafu_Rom})
reduces to the Cauchy distribution). 

In summary, and for all the reasons given above,
the Romanovski polynomials qualify  as the most adequate
real degrees of freedom in the mathematics of the hyperbolic
Scarf potential.

\section{The polynomial construction}
The construction of the $R_n^{\left( a+\frac{1}{2},-2b \right)}(x)$ 
polynomials is now
straightforward and based upon the Rodrigues representation
in eq.~(\ref{Rodrigues}) where we plug in the weight function from
 eq.~(\ref{wafu_Rom}). In carrying out the differentiations we find the
lowest four (unnormalized) polynomials as
\begin{eqnarray}
R^{\left(a+\frac{1}{2}, -2b\right)}_{0}(x)&=&1\, ,
\label{d0}\\
R^{\left(a+\frac{1}{2}, -2b\right)}_{1}(x)&=&-2b +(1-2a)x\, ,
\label{d1}\\
R^{\left(a+\frac{1}{2}, -2b\right)}_{2}(x)&=&3-2a+4b^2-8b(1-a)x +
(6-10a+4a^2)x^2\, ,
\label{d2}\\
R^{(a+\frac{1}{2}, -2b)}_{3}(x)&=&
-266 +12ab -8b^3 +
\lbrack -3(-15 +16a-4a^2) +
12(3-2a)b^2\rbrack x \nonumber\\
&+&(-72b +84ab -24a^2b)x^2
+2(-2+a)(-15 +16a -4a^2)x^3\, 
\, .
\label{d3}
\end{eqnarray}
As illustration, in Fig.~3 we show the Scarf II  wave functions 
of the first and third  levels. 

\vspace{1cm}
\begin{figure}[htb]
\vskip 5.0cm
\includegraphics{psi1.ps}
\includegraphics{psi3.ps}
\vspace{1.01cm}
{\small Fig.~3.
Wave functions for the first and third levels 
 within the hyperbolic Scarf potential.}
\end{figure}
\vspace{0.15cm}

The finite orthogonality of
the Romanovski polynomials becomes  especially transparent
in the interesting limiting case of the $\sech^2 z$ potential
(it appears  in the non-relativistic reduction of the sine-Gordon
equation) where one
easily finds that the normalization constants, $N^{(a+\frac{1}{2}, 0)}_n$,
are given by the following expressions:
\begin{eqnarray}
\left( N^{\left(a+\frac{1}{2}, 0\right)}_1\right)^2&=& \frac{
(2a-1)^2\sqrt{\pi}\Gamma (a-1)}
{2\Gamma(a+\frac{1}{2})}\, , \quad a>1 \, ,\nonumber\\
\left( N^{\left(a+\frac{1}{2}, 0\right)}_2\right)^2&=& 
\frac{2\sqrt{\pi}(a-1)\Gamma (a-2 )}
{\Gamma(a-\frac{1}{2})
} (3-2a)^2,\quad a>2\, ,\nonumber\\
\left( N^{\left(a+\frac{1}{2}, 0\right)}_3\right)^2&=&
\frac{3\sqrt{\pi}(a-2)\Gamma(a-3) }{
\Gamma(a-\frac{1}{2})
}(4a^2-16a +15)^2 ,\quad a>3\,\,\,  {\mbox{etc}}.
\label{norm_4}
\end{eqnarray}
Software like Maple or Mathematica are quite useful
for the graphical study of these functions.
The latter expressions show  that for positive integer values of the $a$
parameter, 
$a =n$, only the first $(n-1)$ Romanovski polynomials are orthogonal
(the convergence of the integrals requires $n<a$),
as it should be in accord with eqs.~(\ref{orth_cond}), (\ref{enrg_hip}).
The general expressions for the normalization constants of any Romanovski
polynomial are defined by integrals of the type
$\int_{-\infty}^{+\infty}(1+x^2)^{-(p-n)}e^{q\tan^{-1}x}dx$
and are analytic for $(p-n)$ integer or semi-integer.

\section{Romanovski polynomials
and non-spherical angular functions in electrodynamics  with non-central 
potentials.}

In recent years there have been several studies of the bound states
of an electron within a compound Coulomb- and a non-central
potential (see Refs.~\cite{Dutt,Kocak} and references therein).
Let us assume the following potential
\begin{equation}
V(r,\theta )=V_{\mathcal C}(r)+\frac{V_2(\theta )}{r^2}\, ,
\quad V_2(\theta )=-c \cot \theta\, ,
\label{non_ctrl_pt}
\end{equation}
where $V_{\mathcal C}(r)$ denotes the Coulomb potential
and $\theta $ is the polar angle (see Fig.~4).
The corresponding Schr\"odinger equation reads 
\begin{equation}\label{SEE-SCC}
\left[
-\left[
\frac{1}{r^{2}}
\frac{\partial}{\partial
r}r^{2}\frac{\partial }{\partial
r}+\frac{1}{r^{2}\sin \theta }\frac{\partial}{\partial\theta}
\sin \theta \frac{\partial }{\partial
\theta}+\frac{1}{r^{2}\sin^{2}\theta}\frac{\partial^{2}
}{\partial\phi^{2}}\right]+V(r,\theta )\right]\Psi(r,\theta ,\varphi )=E
\Psi (r,\theta ,\varphi )\,,
\end{equation}
and is solved as usual by separating variables,
\begin{equation}\label{SSVV}
\Psi(r,\theta,\phi)= {\mathcal R}(r) \Theta (\theta)\Phi ({\varphi})\, .
\end{equation}
The radial and angular differential equations for ${\mathcal R}(r)$ and
$\Theta ( \theta )$ are then found as
\begin{equation}\label{ecuacionRadial}
\frac{{\rm d}^{2}{\mathcal R}(r)}{{\rm d}r^{2}}+\frac{2}{r}
\frac{{\rm d}{\mathcal R}(r)}{{\rm d}r}+\left[
(V_{\mathcal C}(r) + E) -\frac{l(l+1)}{ r^{2}})\right]{\mathcal R}(r)=0,
\end{equation}
and
\begin{equation}\label{ecuacionAngular}
\frac{{\rm d}^{2}\Theta ( \theta )}{{\rm d}\theta ^{2}}+\cot(\theta)
\frac{{\rm d}\Theta  ( \theta )}{{\rm d}\theta}+\left[l(l+1)-
V_2(\theta) -\frac{m^{2}}{\sin^2 \theta }\right]
\Theta  ( \theta )=0\, ,
\end{equation}
with $l(l+1)$ being the separation constant.
{}From now on we will focus attention on eq.~(\ref{ecuacionAngular}).
It is obvious that for $V_2(\theta )=0,$ and upon changing variables from
$\theta $ to $\cos\theta,$ it transforms into
the associated Legendre equation. Correspondingly,
$\Theta (\theta )$ approaches the associated Legendre functions,
\begin{equation}
\Theta (\theta )\stackrel{V_2(\theta )\to 0}{\longrightarrow} 
P_l^m(\cos \theta)\, ,
\label{ass_Leg}
\end{equation}
an observation that will become important below.

\begin{figure}[htbp]
\centerline{\psfig{figure=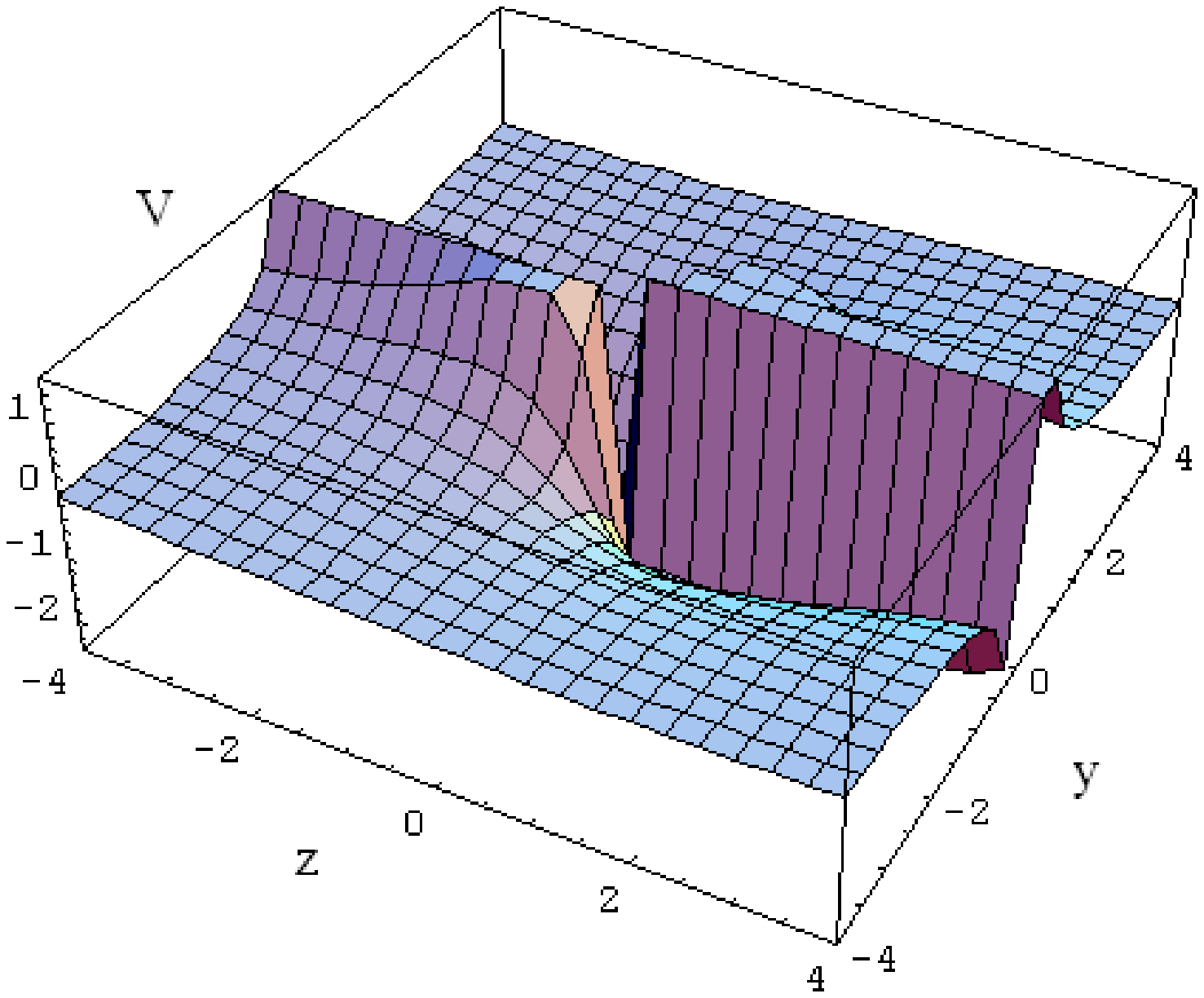,width=7cm}}
\vspace{0.1cm}{\small Fig.~4.
\hspace{0.2cm} 
The non-central potential $V(r,\theta)$,
here displayed in its intersection with the $x=0$ plane,
i.e. for  $r=\sqrt{y^2 +z^2} $, and 
$\theta =\tan^{-1}\frac{y}{z }$. The polar angle part of its exact solutions
is expressed in terms of the Romanovski polynomials.}
\label{fig:brane_string}
\end{figure}

In order to solve eq.~(\ref{ecuacionAngular})
we follow the prescription given in \cite{Dutt} and 
begin by substituting the polar 
angle  by  the new variable, $z$, introduced 
via $\theta \rightarrow f(z)$.
This transformation leads to the new equation
\begin{equation}\label{angular1}
\left[ \frac{{\rm d}^{2}}{{\rm d}z^{2}}+\left[
-\frac{f''(z)}{f'(z)}+
f'(z)\cot f(z)\right]
\frac{{\rm d} }{{\rm d}z }
+\left[
-V_2(f(z))+l(l+1)-
\frac{m^{2}}{\sin^2 f (z)}
\right]
f^\prime\,   ^{2}(z)\right] \psi (z)\ =0,
\end{equation}
with $f'(z)\equiv \frac{{\rm d}f (z)}{{\rm d}z}$, and $\psi (z)$ defined as
 $\psi (z)= \Theta (f(z))$.
Next, one requires 
the coefficient in front of the first derivative to vanish
which transforms eq.~(\ref{angular1}) into a $1d$ Schr\"odinger equation.
This restricts  $f(z)$ to satisfy the differential equation
\begin{equation}\label{angular2}
\frac{f''(z)}{f'(z)}=f'(z)\cot f(z)\, ,
\end{equation}
which is solved by  $f(z)=2\tan^{-1} e^{z}$.
With this relation and after some  algebraic manipulations
one finds that
\begin{equation}
\sin \theta=\frac{1}{\cosh z}, \quad
\cos\theta =-\tanh z\ ,
\label{mapping}
\end{equation}
and consequently,
\begin{equation}
f'(z)=\sin f (z)={\rm sech}\, z .
\label{fprime}
\end{equation}
Equation (\ref{mapping}) implies $\sinh z= -\cot\theta $, or, equivalently,
$\theta =\cot^{-1}(-\sinh z )$.
Upon substituting eq.~(\ref{fprime}) into eqs.~(\ref{non_ctrl_pt}),
and (\ref{angular1}),  one arrives at
\begin{eqnarray}\label{angular3}
\frac{{\rm d}^2 \psi (z)}{{\rm d}z^2 }&+&
\left[
l(l+1)\frac{1}{\cosh^2 z} +
c\tanh z \frac{1}{\cosh z}  - m^{2}\right] \psi (z)=0\,,\nonumber\\
\psi  (z):\stackrel{\mathrm{def}}{=}\Theta \left(
\theta=\cot^{-1}(-\sinh z )\right)\, , &\quad&
\Theta  (\theta ):\stackrel{\mathrm{def}}{=}\psi \left(
z =\sinh^{-1}(-\cot\theta )\right)\,.
\end{eqnarray}
Taking in consideration eqs.~(\ref{Scarf_hip}),(\ref{non_ctrl_pt}), and
(\ref{mapping})
one realizes that the latter equation is precisely the one-dimensional
Schr\"odinger equation with the hyperbolic Scarf potential where
\begin{eqnarray}
l(l+1)=  -(b^2 -a(a+1))\, ,&\quad &c= -b(2a+1),\nonumber\\
m^2 &=& -\epsilon_n =(a-n)^2\, ,\quad m>0\, .
\label{non_sph_1}
\end{eqnarray}
The two parameters of the Romanovski polynomials have to
be determined from the system of the last three equations,
meaning that the $l$, $m$, and $c$ constants
can not be independent. 
There exist various choices for $a$ and $b$.
If defined on the basis of the first two equations, one encounters
\begin{eqnarray}
\left( a+\frac{1}{2}\right)^2&=&
\frac{1}{2}\left(
\left(l+\frac{1}{2}\right)^2
+\sqrt{\left(l+\frac{1}{2}\right)^4 +c^2}\, 
\right),\nonumber\\
b^2&=&
\frac{1}{2}\left(
-\left(l+\frac{1}{2}\right)^2
+\sqrt{\left(l+\frac{1}{2}\right)^4 + c^2}\,
\right)\, .
\label{set_1}
\end{eqnarray}
Substitution of $a$ into the third equation imposes a constraint
on $l$ as a function of $m$, $c$, and $n$.
A second choice for $a$ and $b$ is obtained by 
expressing $a$ from the  third equation in terms of $m$, and $n$
as $a=m+n$ and substituting in the second equation
to obtain $b$ as  
\begin{eqnarray}
b&=&-\frac{c}{2(m+n)+1}.
\label{set_2_b}
\end{eqnarray}
Then the first equation imposes the following restriction on $l$
\begin{eqnarray}
X\stackrel{\mathrm{def}}{:=}(b^2-a(a+1)),  &\quad&
l= -\frac{1}{4} +\sqrt{\frac{1}{4}+X}\, .
\label{parmts}
\end{eqnarray}
This $l$ value which is not necessarily integer, 
is the one that enters the well known energy, 
$E_{n_rl}= -Z^2e^2\mu /(2\hbar^2(n_r+l+1)^2)$, attached to the radial solution,
thus leading to a (discrete) spectrum that no longer bears any resemblance to
the $O(4)$ degeneracy.
This is the path pursued  by Ref.~\cite{Dutt}.
We here instead take a third chance and express 
$a$, $b$, and $c$ as functions of
$l$ alone according to
\begin{eqnarray}
a=b= l(l+1), &\quad& n=a-m=l(l+1)-m,\quad
c=-b(2a+1).
\label{parmts_2}
\end{eqnarray}
This choice allows to consider integer $l$ values.

In making use of  Eqs.~(\ref{p_q_par}),(\ref{bong}),
the solution for $\Theta $  becomes
\begin{eqnarray}
\Theta (\theta )&=& \psi_{n=l(l+1)-m }\left(
z=\sinh^{-1} ( -\cot \theta )\right)\nonumber\\
&=&
(1+\cot^2\theta )^{-\frac{l(l+1)}{2}}
e^{-l(l+1)\tan^{-1} (-\cot \theta )}
R_{l(l+1)-m}^{( l(l+1)+\frac{1}{2},-2l(l+1))}(-\cot\theta ) \, .
\label{non-sph_3}
\end{eqnarray}
The complete angular wave function now can be labeled by $l$ and $m$
(as a tribute to the spherical harmonics) and is given by 
\begin{equation}
Z^{m}_{l}(\theta,\varphi)=\psi_{n=l(l+1)-m}(
\mathrm{\sinh}^{-1}(-\cot(\theta)))e^{im\varphi}.
\end{equation}
It reduces to the spherical harmonics $Y^{m}_{l}(\theta,\varphi)$ 
for $a=b=0$. In this way, the Romanovski polynomials shape the
angular part of the wave function in the problem under consideration.
In the following, we shall refer to $Z_l^m(\theta ,\varphi )$ as
``non-spherical angular functions".  

In Fig.~5 we display two of the lowest $|Z_l^m(\theta ,\varphi )|$ functions
for illustrative purposes. A more extended sampler can be found in 
Ref.~\cite{tesis_DA}.
A comment is in order on  $|Z_l^m(\theta ,\varphi )|$. 
In that regard, it is important to become aware of the fact 
already mentioned above that
the Scarf II potential possesses $su(1,1)$ as a potential algebra,
a result reported by Refs.~\cite{JPA_MG_34,Suk_SU1_1} among others. 
There, it was pointed out that the respective Hamiltonian, $H$, equals 
$H=-C-\frac{1}{4}$, with $C$ being the $su(1,1)$ Casimir operator,
whose eigenvalues in our convention are $j(j-1)$ with $j>0$
versus $j(j+1)$ and $j<0$ in the convention of \cite{JPA_MG_34,Suk_SU1_1}.
As a consequence, the bound state solutions to Scarf II are 
assigned to infinite discrete unitary irreducible
representations, 
$\lbrace D^+_j\, ^ {(m^\prime)}(\theta ,\varphi )\rbrace $, 
of the $SU(1,1)$ group.
The $SU(1,1)$ labels  $m^\prime $, and $j$ are mapped onto ours via
\begin{equation}
m^\prime =a+\frac{1}{2}=l(l+1)+\frac{1}{2},
\quad j=m^\prime -n\,, \quad m^\prime = j,j+1, j+2, ....
\end{equation}
meaning that both $j$ and $m^\prime $ are a half-integer.
The representations are infinite because
for a fixed $j$ value,  $m^\prime$  is bound from below to
$m^\prime _{\mathrm{min}}=j$, but it is not bound from above.

In terms of the $SU(1,1)$ labels the energy rewrites as 
$\epsilon_n=-(j-\frac{1}{2})^2$. The condition $a>n$ translates now as
$j>\frac{1}{2}$. In result,  
$\Theta (\theta )$ becomes
\begin{eqnarray}
\Theta (\theta )= \psi_{n=m^\prime -j}\left(\sinh^{-1}(-\cot \theta)\right)
&=&\sqrt{(1+\cot^2\theta )^{-m^\prime +\frac{1}{2} }
e^{-2b\tan^{-1} (-\cot \theta )}}
R_{m^\prime -j}^{( m^\prime ,-2b)}(-\cot\theta )\nonumber\\ 
&=& D^+ _{j=m+\frac{1}{2}}\, 
^{\left( m^\prime =l(l+1)+\frac{1}{2} \right)}
(\theta ,\varphi)
e^{-im^\prime \varphi }.
\label{su_1_1}
\end{eqnarray}
Here we kept the parameter  $b$ general because its value does not affect
the $SU(1,1)$ symmetry.
Within this context,  $|\psi_ {m^\prime -j}\left(\sinh^{-1}(-\cot\theta 
\right)|$ can be viewed as the absolute value of  a component of a
irreducible $SU(1,1)$ representation   \cite{Wybourne},\cite{Kim-Noz}
realized in terms of the Romanovski polynomials.
The $|Z_l^m(\theta ,\varphi )|$ functions  are then images in 
polar coordinate space of the 
$|D^+
_{j=m+\frac{1}{2}
}\, 
^{
\left(
m^\prime =l(l+1)+\frac{1}{2}
\right)
}|$ components.

\vspace{1truecm}

\begin{figure}[htb]
\vskip 5.0cm
\includegraphics{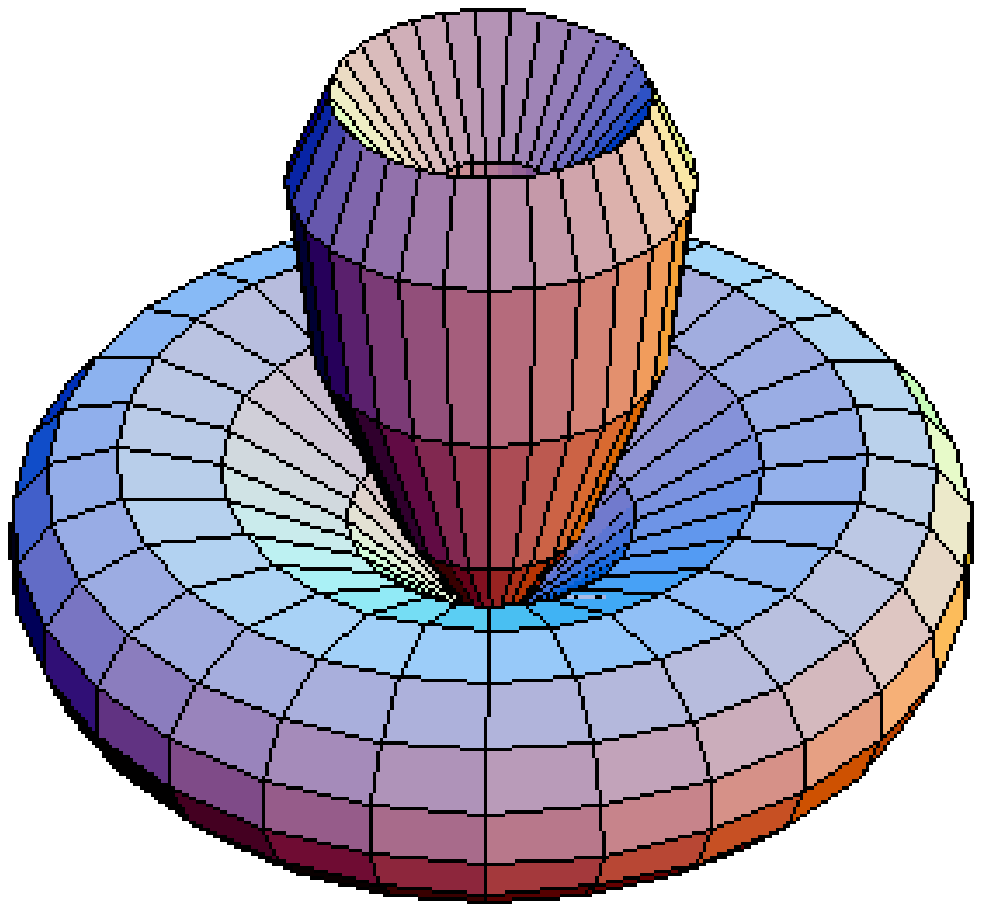}
\includegraphics{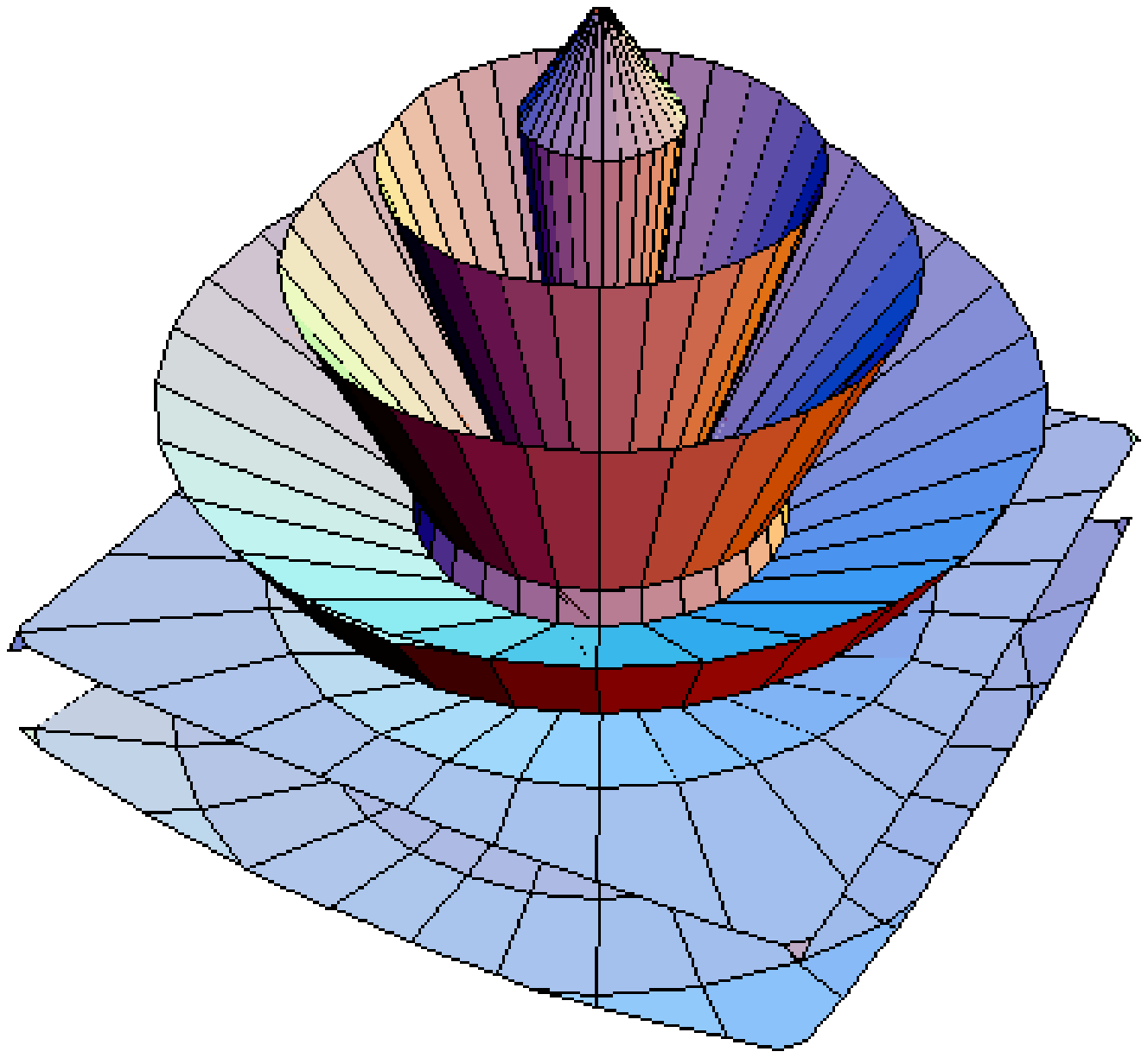}
\vspace{1.01cm}
{\small Fig.~5. Graphical presentation of the 
non-spherical-angular functions  $|Z_1^1(\theta ,\varphi )|$ 
(left), and $|Z_2^1(\theta ,\varphi )|$  (right)  
to the $V_2(r,\theta )$  potential
in eq.~(\ref{non_ctrl_pt}). They portray in polar coordinate space
in turn the components 
$|D^+_{j=\frac{3}{2}}\, ^{\left( m^\prime=\frac{5}{2}\right)} 
(\theta ,\varphi )|$, and 
$|D^+_{j=\frac{3}{2}}\, ^{\left( m^\prime=\frac{13}{2}\right)} 
(\theta ,\varphi )|$
of the respective infinite discrete unitary $SU(1,1)$ representation.}
\end{figure}
\vspace{0.15cm}

\subsection{Romanovski polynomials and associated Legendre functions.} 

It is quite instructive to consider the case of a vanishing
$V_2(\theta )$, i.e. $c=0$, and compare eq.~(\ref{angular3}) to
eq.~(\ref{Scarf_hip}) for $b=0$.
In this case 
\begin{eqnarray}
l=a\, , &\quad & m^2=(l-n)^2\, ,
\end{eqnarray}
which allows one to relate $n$ to $l$ and $m$ as $m=l-n$.
As long as the two equations are equivalent, their solutions differ
at most by a constant factor. This allows to establish a relationship
between the associated Legendre functions and the Scarf II wave functions.
Considering eqs.~(\ref{azero_1}), and (\ref{wfu_x}) together with
eqs.~(\ref{mapping}), one finds
 $\cot\theta = -\sinh z $ which  produces the
following intriguing relationship between the associated Legendre
functions and the Romanovski polynomials 
\begin{equation}
P_l^{m}( \cos\theta  )\sim 
(1+\cot^2 \theta )^{-\frac{l}{2}}
R_{l-m}^{(l+\frac{1}{2}, 0 )}(-\cot \theta)\, , \quad l-m=n=0,1,2,...l.
\label{Ass_Leg_Rom}
\end{equation}
In substituting the latter expression  into the orthogonality integral between
the associated Legendre functions,
\begin{equation}
\int_0^\pi P_l^m(\cos \theta )P_{l^\prime}^m(\cos \theta ){\rm d}\cos \theta\,
=0\, , \quad l\not=l^\prime ,
\label{orth_assc_Leg}
\end{equation}
results in the following integral 
\begin{equation}
\int_0^\pi (1+\cot^2\theta )^{-\frac{l+l^\prime }{2}} 
R_{l-m}^{\left(l+\frac{1}{2}, 0\right)}(-\cot\theta )
R_{l^\prime -m}^{\left(l^\prime +\frac{1}{2}, 0 \right)}
(-\cot\theta )
{\rm d}\cos \theta\,
= 0\, , \quad l\not= l^\prime .
\label{orth_as_Leg}
\end{equation}
Rewriting in conventional notations, the latter expression becomes
\begin{eqnarray}
\int _{-\infty }^{+\infty }
\sqrt{
w^{
\left(l+ \frac{1}{2}, 0 \right) }
(x)} R^
{(l+\frac{1}{2}, 0)}_{n=l-m}(x) \sqrt{
w^{\left( l^\prime + \frac{1}{2}, 0 \right) }
(x)} R^{(l^\prime +\frac{1}{2}, 0 )}_{n^\prime =l^\prime -m }(x ) 
\frac{{\rm d}x}{1+x^2}
&=& 0\, , \quad l\not= l^\prime \,,\nonumber\\
x=\sinh z\, , \quad l-n=l^\prime -n^\prime &=&m\ge 0\, .
\label{Leg_Rom_orth}
\end{eqnarray}
This integral  describes  orthogonality between an {\it infinite\/} 
set of Romanovski
polynomials with  {\it different polynomial parameters\/} (they would
define
wave functions of states bound in {\it different potentials\/}). 
This new orthogonality relationship
does not contradict the finite orthogonality in eq.~(\ref{orth_cond}) which
is valid for states belonging to {\it same potential\/} 
({\it equal polynomial parameters\/}).
Rather, for different potentials eq.~(\ref{orth_cond})
can be fulfilled for an infinite number of states. 
To see this let us consider for simplicity  $n=n^\prime =l-m $, i.e.,
$l=l^\prime$.  Given $p=l+\frac{1}{2}$,
the condition in eq.~(\ref{orth_cond}) defines normalizability and
takes the form
\begin{eqnarray}
2(l-m)<2(l+\frac{1}{2})-1=2l\, ,
\end{eqnarray}
which is automatically fulfilled for any $m>0$. The presence of the
additional factor of $(1+x^2)^{-1}$ guarantees convergence also
for $m=0$.
Equation (\ref{Leg_Rom_orth})
reveals that for parameters attached to the degree of the polynomial, 
an infinite number of  Romanovski polynomials can appear orthogonal,
although not precisely with respect
to the weight function that defines  their Rodrigues
representation.  
The study  presented here is  similar to Ref.~\cite{CK}.
There, the exact solutions of the Schr\"odinger equation with the 
trigonometric Rosen-Morse  potential have been expressed
in terms of Romanovski polynomials (not recognized as such at that time)
and also with parameters that depended on the degree of the polynomial.
Also in this case, the $n$-dependence of the parameters, and the
corresponding varying  weight function allowed to fulfill
eq.~(\ref{orth_cond})  for an infinitely many polynomials. 

\begin{table}
\begin{center}
\begin{tabular}{||l|c|c|c|c|l||}
\hline \hline
Notion & Symbol & $w(x)$  &  Interval & 
Number of orth. polynomials \\
\hline \hline
Jacobi & $P_n^{\nu, \mu }(x)$ & $(1-x)^{\nu}(1+x)^{\mu}$ & 
$[-1,1]$ & infinite 
\\
\hline
Hermite & $ H(x)$ & 
$e^{- x^2}$ &  $
[-\infty,\infty ]$ & infinite \\
\hline \hline
Laguerre & $ L^{\alpha,\beta }(x)$ & 
$x^\beta e^{-\alpha x^2} $ &  $ [0, \infty ]$ & infinite \\
\hline \hline
Romanovski 
& $R_n^{(p ,q )}(x)$ & $(1+x^2)^{-p}\e^{q \tan^{-1} x}$ &  
$[-\infty,\infty ]$ & finite \,\,\\
\hline \hline
\end{tabular}
\vspace{0.5cm}
\begin{flushleft}
Table 1. Characteristics of the orthogonal polynomial 
solutions to the generalized hypergeometric equation. 
\end{flushleft}
\end{center}
\end{table}

\section{Summary}

In this work we presented the classification of the orthogonal polynomial
solutions to the generalized hypergeometric equation in the schemes of
Koepf--Masjed-Jamei \cite{Koepf}, on the one side, 
and Nikiforov-Uvarov \cite{NikUv}, on the other.
We found among them the real polynomials that define the
solutions of the bound states within the hyperbolic Scarf potential.
These so called Romanovski polynomials have the remarkable property that 
for any given set of 
parameters, only a finite number of them is orthogonal.
In such a manner, the finite number of bound states within Scarf II 
were mapped  onto a finite set of orthogonal polynomials of a new type.
 
We showed that the Romanovski polynomials define also the
angular part of the wave function of the non-central potential
considered in section IV. Yet, in this case,the  polynomial parameters
resulted dependent on the polynomial degree.
We identified these non-spherical angular solutions
to the non-central potential under investigation
as images in polar coordinate space of components of infinite
discrete unitary $SU(1,1)$ representations.
In the limit of the vanishing non-central piece of the potential,
we established a non-linear relationship between the Romanovski polynomials
and the associated Legendre functions. On the basis of the orthogonality
integral for the latter we   derived a new
such integral for the former.

The  presentation contains all the details which in our understanding 
are indispensable for reproducing our results.
With that we worked out two problems  which could be
used in the class on quantum mechanics and on 
mathematical methods in physics as well and which allow to
practice performing  with symbolic softwares.
The appeal of the two examples is that they
simultaneously  relate to relevant peer research.

The hyperbolic Scarf potential and its exact solutions
are interesting mathematical entities on their own, with
several applications in physics, ranging from super-symmetric 
quantum mechanics over soliton physics to field theory. 
We expect future research to reveal more 
and interesting properties and problems related
to the hyperbolic Scarf potential and its 
exact real polynomial solutions.

\section*{Acknowledgments}
We are indebted to  Dr.\ Jose-Luis Lop\'ez Bonilla 
for bringing the important references ~\cite{Zarzo}, and
\cite{Kocak} to our attention.
We furthermore thank Drs. Hans-J\"urgen Weber and Alvaro P\'erez Raposo for 
insightful discussions on the orthogonality
issue.  We benefited from the lectures on supersymmetric quantum mechanics
at the XXXV  Latin American School (ELAF), ``
Super-symmetries in Physics and Applications'',
held in M\'exico in the Summer of 2004.

Work supported by Consejo Nacional de Ciencia y 
Technolog\'ia (CONACyT) Mexico under grant number C01-39280.

\end{document}